\documentclass[prc,superscriptaddress,twocolumn,showpacs,preprintnumbers]{revtex4}
\usepackage[dvips]{graphicx}
\usepackage{amsmath}
\usepackage{amssymb}
\usepackage{ulem}
\usepackage{times}
\usepackage{mathrsfs}
\usepackage{multirow}
\usepackage{bm}

\newcommand{\beq}{\begin{equation}}
\newcommand{\eeq}{\end{equation}}
\newcommand{\bea}{\begin{eqnarray}}
\newcommand{\eea}{\end{eqnarray}}

\newcommand{\bfi}[1]{\mbox{\boldmath $#1$}}

\newcommand{\vb}{{\bfi b}}

\newcommand{\vR}{{\bfi R}}

\def\mbold#1{\mbox{\boldmath $#1$}}

\def\beq{\begin{equation}}
\def\eeq{\end{equation}}
\def\bea{\begin{eqnarray}}
\def\eea{\end{eqnarray}}
\def\la{\mathrel{\mathpalette\fun <}}

\def\fun#1#2{\lower3.6pt\vbox{\baselineskip0pt\lineskip.9pt
  \ialign{$\mathsurround=0pt#1\hfil##\hfil$\crcr#2\crcr\sim\crcr}}}
\def\mbold#1{\mbox{\boldmath $#1$}}

\begin{document}

\title{Reaction mechanism in odd-even staggering of reaction cross sections
}

\author{Satoru Sasabe}
\email[]{sasabe@email.phys.kyushu-u.ac.jp}
\affiliation{Department of Physics, Kyushu University, Fukuoka 812-8581, Japan}

\author{Takuma Matsumoto}
%\email[]{matsumoto@phys.kyushu-u.ac.jp}
\affiliation{Department of Physics, Kyushu University, Fukuoka 812-8581, Japan}

\author{Shingo Tagami}
%\email[]{tagami@phys.kyushu-u.ac.jp}
\affiliation{Department of Physics, Kyushu University, Fukuoka 812-8581, Japan}

\author{Naoya Furutachi}
%\email[]{furutati@nucl.sci.hokudai.ac.jp}
\affiliation{Department of Physics, Hokkaido University, 
Sapporo 060-0810, Japan}

\author{Kosho Minomo}
%\email[]{minomo@phys.kyushu-u.ac.jp}
\affiliation{Department of Physics, Kyushu University, Fukuoka 812-8581, Japan}

\author{Yoshifumi R. Shimizu}
%\email[]{shimizu@phys.kyushu-u.ac.jp}
\affiliation{Department of Physics, Kyushu University, Fukuoka 812-8581, Japan}

\author{Masanobu Yahiro}
%\email[]{yahiro@phys.kyushu-u.ac.jp}
\affiliation{Department of Physics, Kyushu University, Fukuoka 812-8581, Japan}

\date{\today}

\begin{abstract}
It was recently suggested that the odd-even staggering of reaction cross sections is an evidence of 
the pairing anti-halo effect on projectile radii.  
We define the dimensionless staggering parameters, $\Gamma_{\rm rds}$ and $\Gamma_{\rm R}$, 
for projectile radii and reaction cross sections, respectively, and analyze the relation between 
$\Gamma_{\rm rds}$ and $\Gamma_{\rm R}$ for the scattering of $^{14,15,16}$C 
from a $^{12}$C target at 83~MeV/A  by taking account of projectile-breakup and nuclear-medium effects newly with 
the microscopic version of  the continuum discretized coupled-channels method. 
The value of $\Gamma_{\rm R}$ is deviated from that of $\Gamma_{\rm rds}$ 
by the projectile-breakup effect, the nuclear-medium effect and 
an effect due to the fact that the scattering are not 
the black-sphere scattering (BSS) exactly. 
The projectile-breakup and nuclear medium effects are nearly canceled 
for $\Gamma_{\rm R}$. 
The remaining non-BSS effect becomes small as an incident energy decreases, indicating that nucleus-nucleus scattering at lower incident energies are a good probe of evaluating $\Gamma_{\rm rds}$ from measured reaction cross sections. 
\end{abstract}


\maketitle

{\it Introduction.} 
Interaction cross section $\sigma_{\rm I}$ and reaction cross section $\sigma_{\rm R}$ are an important tool of 
determining radii of unstable nuclei. Actually, 
the halo structure as an exotic property was reported 
for unstable nuclei like $^{11}$Li  through 
analyses of measured $\sigma_{\rm I}$~\cite{Tanihata85,Ozawa01}. 
Very recently,  $\sigma_{\rm I}$ was measured for Ne isotopes~\cite{Takechi12} and it is suggested by the analyses~\cite{Minomo:2011sj,Minomo:2011bb} that $^{31}$Ne 
is a halo nucleus with large deformation.

The difference between $\sigma_{\rm I}$ and $\sigma_{\rm R}$ is considered to be small 
for scattering of unstable nuclei at intermediate energies~\cite{Sumi:2012fr}. 
The reaction cross section is nearly proportional to a raidus of projectile; for example, see Ref.~\cite{Sumi:2012fr} for detailed analyses. 
Meanwhile, it is well known that pairing correlations are important particularly in even-$N$ nuclei. 
The correlations become essential in weakly bound nuclei, since 
they are not bound without the correlations. 
Effects of the correlations on nuclear radii of unstable nuclei were investigated by the Hartree-Fock Bogoliubov (HFB) method~\cite{Bennaceur00}. 
In the mean-field picture, the correlations make the quasi-particle energy larger and hence reduce the root-mean-square radius of 
the HFB density. Obviously, this effect is conspicuous for 
unstable nuclei with the separation energy smaller than the gap energy.   
Thus, the pairing correlation suppresses the growth 
of halo structure for even-even unstable nuclei. 
This is now called the pairing anti-halo effect. 

The pairing anti-halo effect is an interesting phenomenon, but any clear evidence is not shown for the effect yet. 
Very recently, however, Hagino and Sagawa suggested that 
observed odd-even staggerings of $\sigma_{R}$ are possible  
evidence of the effect~\cite{Hagino:2011ji,Hagino:2011aa,Hagino:2012qu}. 
They introduced the staggering parameter \cite{Hagino:2012qu} 
\bea
\gamma_3=-\frac{\sigma_{\rm R}(A)-2\sigma_{\rm R}(A+1)+\sigma_{\rm R}(A+2)}{2},
\eea
where the mass number $A$ of projectile is assumed to be even.  
In Ref.~\cite{Hagino:2011ji}, the staggering was analyzed with the HFB method for $^{30,31,32}$Ne+$^{12}$C scattering at 240~MeV/A~\cite{Takechi12}  and with the three-body model for $^{14,15,16}$C+$^{12}$C scattering at 83~MeV/A~\cite{Fang04}. 
The analyses are 
successful in reproducing observed staggerings~\cite{Takechi12,Fang04}, 
although the reaction calculations are based on the Glauber model. 

In this paper, we reanalyze not $^{30,31,32}$Ne but $^{14,15,16}$C scattering in order to focus our attention on the reaction mechanism, since $^{15}$C has a simpler structure than $^{31}$Ne in the sense that the first excited energy of $^{14}$C as a core nucleus is much larger than that of $^{30}$Ne. For $^{14,15,16}$C, $\gamma_3$ is 
163 $\pm$ 52 mb and about 10 \% of 
$\sigma_{\rm R}({^{15}}{\rm C})=1319 \pm 40$ mb~\cite{Fang04}.  
Thus the pairing anti-halo effect may be comparable with the projectile-breakup and nuclear-medium effects that are not taken into account in the previous analysis. 
Therefore, we investigate these effects on the staggering, using 
the continuum-discretized coupled-channels method
(CDCC)~\cite{Kam86,Aus87,Yah12}. 
CDCC for two-body (three-body) projectiles is often called 
three-body (four-body) CDCC; in the naming the target degree of freedom 
is taken into account. This is the first application of four-body CDCC to $^{16}$C.

{\it Theoretical framework.} 
Following Ref.~\cite{Hagino:2011ji}, we assume the $n$ + $^{14}$C
two-body model for $^{15}$C and the $n$ + $n$ + $^{14}$C three-body model 
for $^{16}$C. The three-body model of $^{16}$C is a simple
model for treating pairing correlations between extra two neutrons.  
%Scattering of $^{16}$C from  $^{12}$C is then described by 
%the $n$ + $n$ + $^{14}$C + $^{12}$C four-body model. 
%The scattering is governed by the Schr\"{o}dinger equation
In the present calculation, breakup reactions of $^{15}$C and $^{16}$C
on $^{12}$C are described by the $n$ + $^{14}$C + $^{12}$C three-body
model and the $n$ + $n$ + $^{14}$C + $^{12}$C four-body model,
respectively. The Schr\"{o}dinger equation is defined as
\begin{eqnarray}
(H-E)\Psi=0
\label{original-H}
\end{eqnarray}
for the total wave function $\Psi$, where $E$ is an energy of the total
system. The total Hamiltonian $H$ is  defined by  
\bea
 H=K_{R}+U+h, 
\label{H4}
\eea
where $h$ denotes the internal Hamiltonian of $^{15}$C or $^{16}$C, 
${\mbold R}$ is the center-of-mass coordinate of the projectile 
relative to a $^{12}$C target. The kinetic energy operator associated
with ${\mbold R}$ is represented by $K_{R}$, and $U$ is the sum 
of interaction between the constituents in the projectile and the
target defined as
\bea
U=U_{n}(R_{n})+ U_{^{14}{\rm C}}(R_{^{14}{\rm C}})
+\frac{e^2Z_{\rm P}Z_{\rm T}}{R}, 
\label{potU1}
\eea
for $^{15}$C and
\bea
U=U_{n_1}(R_{n_1})+U_{n_2}(R_{n_2})+ U_{^{14}{\rm C}}(R_{^{14}{\rm C}})
+\frac{e^2Z_{\rm P}Z_{\rm T}}{R}
\label{potU}
\eea
for $^{16}$C, where $U_{x}$ is the nuclear part of the optical potential
between $x$ and $^{12}$C as a function of the relative coordinate
$R_{x}$.  

The optical potential $U_{x}$ %between $x$ and a $^{12}$C target 
is constructed microscopically by folding  
the Melbourne $g$-matrix nucleon-nucleon interaction ~\cite{Amos} 
with densities of $x$ and $^{12}$C. 
For $^{12}$C, the proton density is obtained phenomenologically  from 
the the electron scattering~\cite{C12-density}, and 
the neutron density is assumed to be the same as the proton one, since
the proton root-mean-squared (RMS) radius deviates from 
the neutron one only by less than 1\% in the HFB calculation. 
For $^{14}$C, the matter density is determined by the HFB calculation 
with the Gogny-D1S interaction~\cite{GognyD1S}, 
where the center-of-mass correction is made in the standard
manner~\cite{Sumi:2012fr}. 
As shown latter, the total
reaction cross section calculated with the folding $^{14}$C-$^{12}$C
potential $U_{^{14}{\rm C}}$ is good agreement with the experimental
data for the $^{14}$C + $^{12}$C scattering at 83 MeV/A.
The  Melbourne $g$-matrix folding method is successful in reproducing 
nucleon-nucleus and nucleus-nucleus elastic scattering 
systematically~\cite{Sumi:2012fr,Yah12}. 
The folding potentials thus obtained include 
{\it the nuclear-medium effect}.  
CDCC with these microscopic potentials is the  microscopic version of
CDCC.

In the present system, Coulomb breakup is quite small, since 
the projectile (P) and the target (T) are light nuclei and hence 
the Coulomb barrier energy between P and T is much smaller than 
the incident energy considered here. 
We then neglect Coulomb breakup, as shown in Eq.~\eqref{potU}, 
where $Z_{\rm P}$ and $Z_{\rm T}$  are the atomic numbers of nuclei P
and A, respectively.  

The $n$-$^{14}$C interaction in $h$ of $^{15}$C is taken as the same
interaction as in Ref.~\cite{Hagino:2011ji}, which well reproduce the
properties of ground and 1st-excited states in $^{15}$C.
For the $n$ + $n$ + $^{14}$C system, 
we use the Bonn-A interaction~\cite{Mac89} between two neutrons 
and take the same $n$-$^{14}$C interaction as mentioned above. 
Furthermore the effective three-body interaction is introduced to
reproduce the measured binding energy of $^{16}$C. 
Eigenstates of $h$ are obtained with numerical techniques of
Ref.~\cite{Mat06}, that is, the orthogonality condition is imposed.
%For the $n$ + $^{14}$C subsystem, the present $h$ well 
%reproduces properties of $^{15}$C in its its ground state and 
%first-excited state.  
%The ground state of $^{14}$C obtained by the deformed HFB calculation 
%is spherical. 
Now we introduce the dimensionless staggering parameter 
$\Gamma_{\rm rds}$ for the RMS radii ${\bar r}$ of P and T:   
\bea
\Gamma_{\rm rds} = 
\frac{{\bar R}^2(A+1) - [{\bar R}^2(A) + {\bar R}^2(A+2)]/2}
{[{\bar R}^2(A+2) - {\bar R}^2(A)]/2}
\eea
with
\bea
{\bar R}(A) = {\bar r}(A) + {\bar r}(T).
\eea
Here 
$\Gamma_{\rm rds} \ge 1$ when ${\bar r}(A+1) \ge {\bar r}(A+2)$.
%and 0 when ${\bar r}^2(A+1)=[{\bar r}^2(A)+{\bar r}^2(A+2)]/2$. 
Matter radii of $^{14,15,16}$C are summarized in Table \ref{Table}.  
The present two- and three-model yields 
$\Gamma_{\rm rds}=1.3$ for $^{14,15,16}$C.

\begin{table}[htbp]
\caption{Matter radii of $^{14,15,16}$C.  
}
\begin{tabular}{c|ccc}
\hline \hline
\makebox[45pt][c]{} & \makebox[45pt][c]{${\bar r}(^{14}$C) [fm]} &
 \makebox[45pt][c]{${\bar r}(^{15}$C) [fm]} 
& \makebox[45pt][c]{${\bar r}(^{16}$C) [fm]} \\
\hline
\makebox[45pt][c]{Calc.} & 2.51\footnote{Present calculation.} 
     & 2.87\footnotemark[1] & 2.83\footnotemark[1]\\
\makebox[45pt][c]{} & 2.53\footnote{Ref.~\cite{Hagino:2011ji}.} & 
	 2.90\footnotemark[2] & 
	     2.81\footnotemark[2]\\
\makebox[45pt][c]{Exp.} & 2.50\footnote{Charge radius~\cite{Sc82e}.} & - & -\\
%\hline
% $^{14}$C & 2.51 \\
% $^{15}$C & 2.87 \\
% $^{16}$C & 2.83 \\
\hline \hline
\end{tabular}
\label{Table}
\end{table}

In CDCC, eigenstates of $h$ consist of finite number of discrete states 
with negative energies and discretized-continuum states with positive
energies. The Schr\"{o}dinger equation \eqref{original-H} is 
solved in a modelspace ${\cal P}$ spanned 
by the discrete and discretized-continuum states:
\begin{eqnarray}
{\cal P}(H-E){\cal P}\Psi_{\rm CDCC}=0 .
\label{H-CDCC}
\end{eqnarray} 
Following Ref.~\cite{Mat03}, 
we obtain the discrete and discretized continuum states 
by diagonalizing $h$ in a space spanned by the Gaussian basis
functions. 
This discretization is called the pseudo-state method.
The elastic and discrete breakup $S$-matrix elements are
obtained by solving the CDCC equation \eqref{H-CDCC} under the standard
asymptotic boundary condition~\cite{Kam86,Piy89}. In actual
calculations, we neglect the projectile spin, since the effect on
$\sigma_{\rm R}$ is small~\cite{Hashimoto:2011nc,Sumi:2012fr}.  
We take the angular momentum between $n$ and $^{14}$C
for breakup states of $^{15}$C up to g-wave, and $0^+$ and $2^+$
breakup states of $^{16}$C.

Now we define the dimensionless staggering parameter also for 
$\sigma_{\rm R}$:
\bea
\Gamma_{\rm R}&=&\frac{\gamma_3}{[\sigma_{\rm R}(A+2)-\sigma_{\rm R}(A)]/2},  
\label{OES-para-2}
\eea
where $\Gamma_{\rm R}=0$ when 
$\sigma_{\rm R}(A+1)=[\sigma_{\rm R}(A+2)+\sigma_{\rm R}(A)]/2$ and 
$\Gamma_{\rm R}=1$ when $\sigma_{\rm R}(A+1)=\sigma_{\rm R}(A+2)$. 
When the absolute value of the elastic $S$-matrix element, $|S_{\rm el}(L)|$, 
is 0 for orbital angular momenta $L$ corresponding to 
the nuclear interior and 1 for those to the nuclear exterior, 
it is satisfied that 
$\sigma_{\rm R}(A) \propto {\bar R}^2(A)$
%for the RMS radius ${\bar r}(T)$ of T
~\cite{Hashimoto:2011nc}. 
In the black-sphere scattering, Eq.~\eqref{OES-para-2} is reduced 
to $\Gamma_{\rm R} = \Gamma_{\rm rds}$. 
Once this condition is satisfied, $\Gamma_{\rm R}$ does not depend on 
an incident energy $E_{\rm in}$.  
Three types of models are considered to investigate the nuclear-medium 
and projectile-breakup effects on $\sigma_{\rm R}$. 
\begin{itemize}
\item Model I is the $T$-matrix single-folding model 
that has no nuclear-medium 
and projectile-breakup effects. The $U_x$ are constructed from 
the Melbourne $g$-matrix nucleon-nucleon interaction at zero density. 
The single-channel calculation is done in \eqref{H-CDCC}.   
\item Model II is the $g$-matrix single-folding model that has the nuclear-medium effect but not the projectile-breakup effect. 
This is the same as Model I, but the density dependence of the Melbourne $g$-matrix  is properly taken. 
\item Medel III is the model that has both the nuclear-medium and 
the projectile-breakup effect. CDCC calculations are done for $^{15,16}$C scattering, but 
the $g$-matrix single-folding model is taken for $^{14}$C scattering, 
since $^{14}$C is a tightly-bound system.
\end{itemize}

\begin{figure}[htbp]
\includegraphics[width=0.4\textwidth,clip]{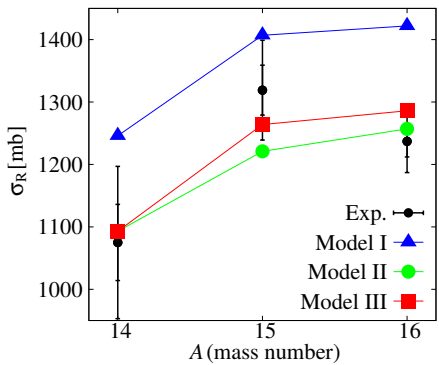}
\caption{(Color online)
Reaction cross sections $\sigma_{\rm R}$ for $^{14,15,16}$C+$^{12}$C 
scattering at 83 MeV/A. 
Triangle, circle and square symbols  
stand for results of Model I, II and III, respectively. 
The experimental data are taken from Ref.~\cite{Fang04}.
}
\label{RCS}
\end{figure}

{\it Results.}
Figure~\ref{RCS} shows $\sigma_{\rm R}$ for $^{14,15,16}$C+$^{12}$C scattering at 83 MeV/A. 
Triangle, circle and square symbols stand for the results of Model I, II and III, respectively. 
Model III well reproduces the experimental data~\cite{Fang04}, whereas
Model I largely overestimates them; here the data are plotted with 2-$\sigma$ error (95.4\% certainty).
The nuclear-medium and projectile-breakup effects are thus important 
for $\sigma_{\rm R}$. 
Model III yields $\Gamma_{\rm R}=0.77$ that is deviated from $\Gamma_{\rm rds}=1.3$.  
When the breakup effect is switched off from Model III, $\sigma_{\rm R}$ is reduced from squares to circles. 
This reduction is most significant for $^{15}$C,  so that $\Gamma_{\rm R}$ is reduced from 0.77 to 0.56. 
Furthermore, when the medium effect is switched off from Model II, 
the $\sigma_{\rm R}$ are enhanced by about 10\% from circles to triangles for all the cases of $^{14,15,16}$C. 
More precisely, the enhancement is 13\% for $^{14,16}$C but 15\% for $^{15}$C, and consequently, 
$\Gamma_{\rm R}$ increases from 0.56 to 0.83 by neglecting 
the medium effect. 
Thus the breakup and medium effects are nearly canceled for $\Gamma_{\rm R}$. 
The resultant value $\Gamma_{\rm R}=0.83$ is still considerably deviated from $\Gamma_{\rm rds}=1.3$. 
This means that the present scattering are not the black-sphere scattering (BSS) exactly.  
This effect is referred to as ``non-BSS effect'' in this paper and is explicitly investigated below.

%%%%%%%%%%%%%%%%%%%%%%%
%%%  Figure 1
%%%%%%%%%%%%%%%%%%%%%%%

\begin{figure}[htbp]
\includegraphics[width=0.4\textwidth,clip]{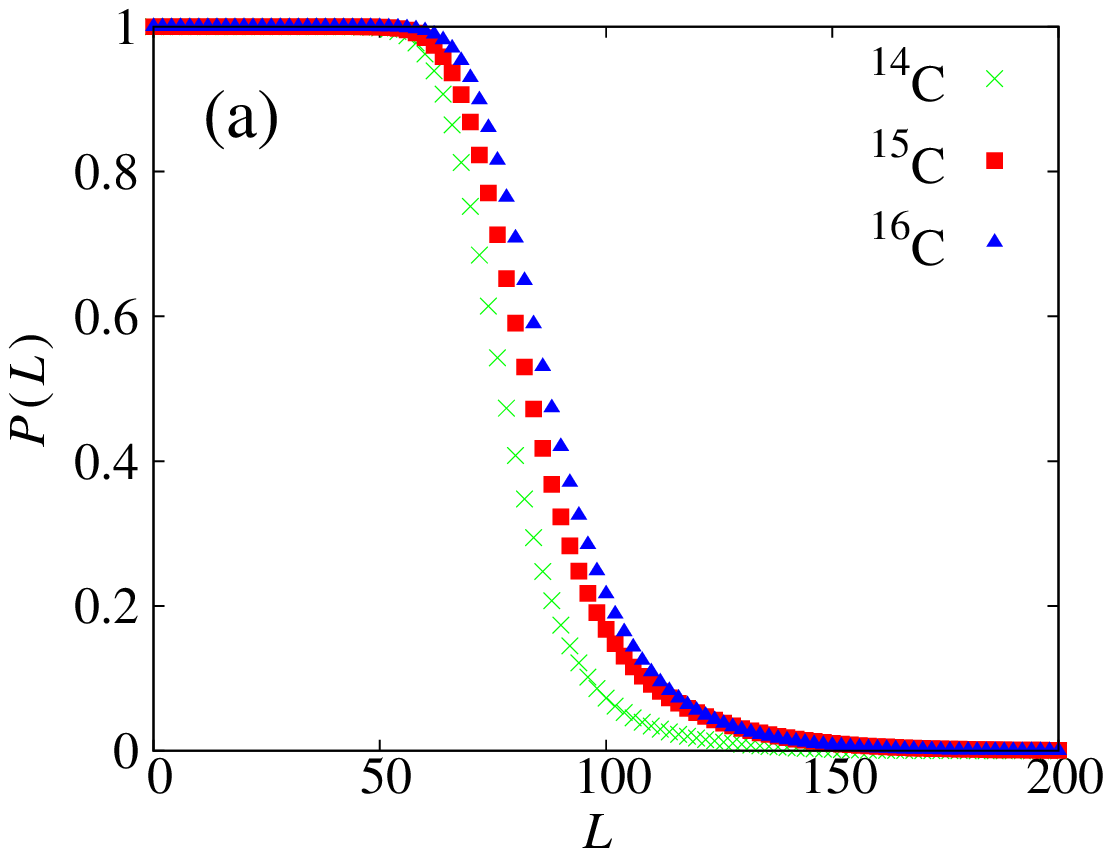}
\includegraphics[width=0.4\textwidth,clip]{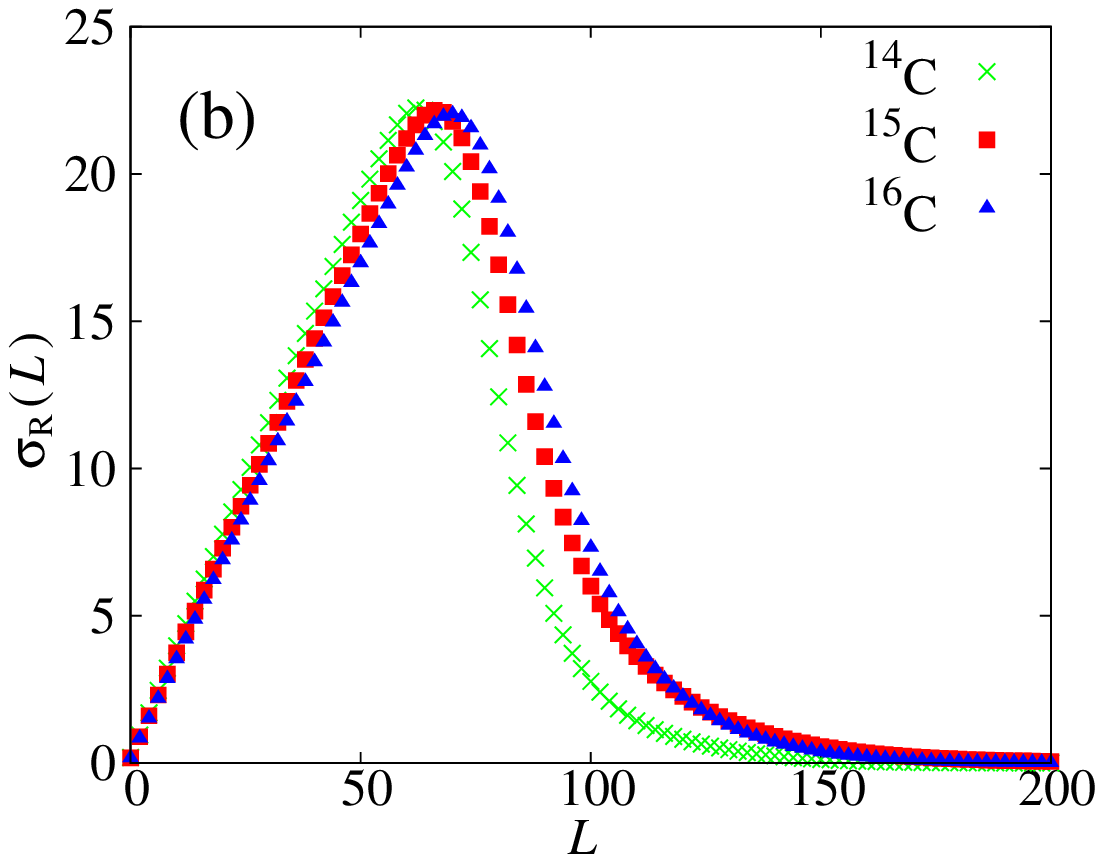}
\caption{(Color online)
$L$ dependence of (a) the absorption probability $P(L)$ and (b) the partial reaction cross section 
for $^{14,15,16}$C+$^{12}$C scattering at 83 MeV/A. Model I is taken. 
 }
\label{PRCS-Model-I}
 \end{figure}

Figure \ref{PRCS-Model-I} shows the absorption probability $P(L) \equiv 1-|S_{\rm el}(L)^2|$ and the partial reaction cross section 
$\sigma_{\rm R}(L) \equiv  (2L+1)P(L)\pi/K^2$ as a function of $L$, where $\hbar K$ is an initial momentum of the 
elastic scattering.  Here Model I is taken. 
For all the $^{14,15,16}$C scattering, $P(L)$ behaves as not a step function but a logistic function. 
Thus the scattering are not the BSS exactly. Furthermore, $L$ dependences of the $P(L)$ are different among the three projectiles 
at $ 60 \la L \la 150$ corresponding to the peripheral region of a $^{12}$C target. As a consequence of the difference, 
$\sigma_{\rm R}$ is not proportional to ${\bar R}^{2}$ properly. 
In fact, $^{15}$C has a larger RMS radius than $^{16}$C, but $^{15}$C scattering has a smaller $\sigma_{\rm R}(L)$ than $^{16}$C one 
at  $ 70 \la L \la 120$ because of the fact that the volume integral of 
the imaginary part of the single-folding potential  $\langle \varphi_{0} | U | \varphi_{0} \rangle$
is smaller for $^{15}$C projectile than for $^{16}$C projectile; here $\varphi_{0}$ is the projectile ground-state wave function.

%%%%%%%%%%%%%%%%%%%%%%%
%%%  Figure 2
%%%%%%%%%%%%%%%%%%%%%%%

\begin{figure}[htbp]
\includegraphics[width=0.4\textwidth,clip]{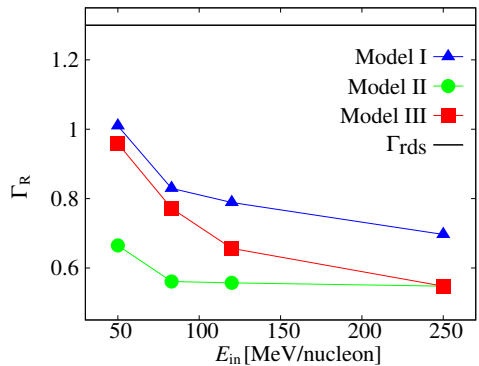}
\caption{(Color online)
$E_{\rm in}$-dependence of $\Gamma_{\rm R}$. 
Triangle, circle and square symbols stand for the results 
of Model I, II and III, respectively.  
At $E_{\rm in}=250$MeV/A, the breakup effect is 
found to be negligible in the previous work~\cite{Sumi:2012fr}, 
so the result of Model III is identified with that of Model II there. 
The solid straight line denotes  $\Gamma_{\rm rds}$.
 }
\label{E-dep}
 \end{figure}

Figure \ref{E-dep} shows $E_{\rm in}$-dependence of $\Gamma_{\rm R}$. 
Triangle, circle and square symbols correspond to the resutls of Model I, II and III, respectively, 
whereas the solid straight line denotes  $\Gamma_{\rm rds}$. 
The deviation of triangles from the solid straight line shows the non-BSS effect,  
the deviation of circles from triangles does the nuclear-medium effect, and 
the deviation of squares from circles comes from the projectile-breakup effect. 
As $E_{\rm in}$ goes up, the breakup effect decreases rapidly, but the non-BSS effect increases. 
The nuclear-medium effect also decreases but very slowly. Thus the non-BSS and medium effects are important for $\Gamma_{\rm R}$ at higher $E_{\rm in}$ around 250~MeV/A. 
At lower $E_{\rm in}$ from 50 to 80 ~MeV/A, meanwhile, 
the medium and breakup effects are nearly canceled, so that 
the non-BSS effect becomes most significant for $\Gamma_{\rm R}$. 
Since the non-BSS effect is smaller at lower $E_{\rm in}$, we can 
conclude  that lower-incident energy scattering are a good probe of evaluating  
$\Gamma_{\rm rds}$ from $\sigma_{\rm R}$.

As mentioned above, the non-BSS effect becomes large as $E_{\rm in}$. 
This can be understood as follows. 
In the high $E_{\rm in}$ where the eikonal approximation is valid, 
$\sigma_{\rm R}$ is proportional to the volume integral of 
the imaginary part $\langle \varphi_{0} | W | \varphi_{0} \rangle$ 
of $\langle \varphi_{0} | U | \varphi_{0} \rangle$~\cite{Yahiro:2011bu,Yah12}, since
\bea
\sigma_{\rm R}&=&
\int d^2 \vb [1-| \langle \varphi_{0} | S | \varphi_{0} \rangle|^2]  
\nonumber \\ 
%&=&
%\int d^2 \vb [1- \langle \varphi_{0} | |S|^2 | \varphi_{0} \rangle]  
%\nonumber \\ 
 &=& 
\frac {-2}{\hbar v_0}\int d^3 \vR  \langle \varphi_{0} | W | \varphi_{0} \rangle
\label{ERT-sigma-R}
\eea
with 
 \bea
  S
&=&
    \exp\Big[   - \frac {i}{\hbar v_0} \int_{-\infty}^{\infty} dZ 
    U \Big] ,
  \label{S} 
\eea
where $v_0$ is the incident velocity of P and $\vR=(\vb,Z)$. 
Equation \eqref{ERT-sigma-R} shows that $\sigma_{\rm R}(A+2)-\sigma_{\rm R}(A)=2(\sigma_{\rm R}(A+1)-\sigma_{\rm R}(A))$ 
and hence $\Gamma_{\rm R}=0$.

%%%%%%%%%%%%%%%%%%%%%%%
%%%  Figure 3
%%%%%%%%%%%%%%%%%%%%%%%

{\it Summary.} 
The present microscopic version of three- and four-body CDCC calculations reproduces  $\sigma_{\rm R}$ for $^{14,15,16}$C+$^{12}$C scattering at 83~MeV/A. 
The projectile-breakup effect is significant for $^{15}$C scattering and appreciable for $^{16}$C scattering, whereas the nuclear-medium effect is sizable for all the $^{14,15,16}$C scattering. 
In general, the $\sigma_{\rm R}$-staggering $\Gamma_{\rm R}$ is
deviated from the radius-staggering $\Gamma_{\rm rds}$ 
by the non-BSS, nuclear-medium and projectile-breakup effects. 
At lower $E_{\rm in}$ from 50 to 80 MeV/A, 
the breakup and medium effects are nearly canceled and 
the remaining non-BSS effect is rather small for $\Gamma_{\rm R}$. Therefore, the lower-$E_{\rm in}$ scattering are 
a good probe of evaluating $\Gamma_{\rm rds}$ from $\sigma_{\rm R}$. 
At high $E_{\rm in}$, meanwhile, the non-BSS effect is significant, whereas  the nuclear-medium and projectile-breakup effects are small or negligible. 
The non-BSS effect largely reduces $\Gamma_{\rm R}$ from  $\Gamma_{\rm rds}$. Thus the radius-staggering $\Gamma_{\rm rds}$ is 
masked by the non-BSS effect at high $E_{\rm in}$. 
This means that if experimental data show a large value of 
$\Gamma_{\rm R}$, the corresponding radius-staggering 
$\Gamma_{\rm rds}$ is even large. 
A good example is the $\sigma_{\rm R}$-staggering for 
$^{30,31,32}$Ne scattering at 250 MeV/A. 
Thus $\Gamma_{\rm R}$ is a good quantity to find exotic properties 
of unstable nuclei. 

%%%%%%%%%%%%%%%%%%%%%%%%%%%%%%%%%%%%%%%%%%%%%%%%%%%%%%%%%%%%%%%%%%%
%                         Acknowledgments                         %
%%%%%%%%%%%%%%%%%%%%%%%%%%%%%%%%%%%%%%%%%%%%%%%%%%%%%%%%%%%%%%%%%%%

The authors would like to thank Fukuda and Yamaguchi 
for helpful discussions. 
This work has been supported in part by the Grants-in-Aid for
Scientific Research of Monbukagakusho of Japan and JSPS.

%%%%%%%%%%%%%%%%%%%%%%%%%%%%%%%%%%%%%%%%%%%%%%%%%%%%%%%%%%%%%%%%%%%
%                          References                             %
%%%%%%%%%%%%%%%%%%%%%%%%%%%%%%%%%%%%%%%%%%%%%%%%%%%%%%%%%%%%%%%%%%%

%\bibliographystyle{h-physrev}
%\bibliography{bib-TM1}

\end{document}